\documentclass[11pt,a4paper]{article} 
\usepackage{jcappub}
\usepackage{natbib}
\usepackage{hyperref}

\title{Super Heavy Dark Matter in light of BICEP2, Planck and Ultra High Energy Cosmic Rays Observations}

\author{R. Aloisio$^{1,2}$, S. Matarrese$^{3,1}$ and A.V. Olinto$^{4}$}

\affiliation{$^{1}$Gran Sasso Science Institute (INFN), L'Aquila,  Italy.\\ 
$^{2}$INAF/Osservatorio Astrofisico di Arcetri, Firenze, Italy.\\ 
$^{3}$Dipartimento di Fisica e Astronomia G. Galilei and INFN, Universit\`a degli Studi di Padova, Padova, Italy.\\
$^{4}$Department of Astronomy \& Astrophysics, Kavli Institute for Cosmological Physics,
The University of Chicago, Chicago (IL), USA.}

\emailAdd{aloisio@arcetri.astro.it, sabino.matarrese@pd.infn.it, olinto@kicp.uchicago.edu}

\abstract{ \\
The announcement by BICEP2 of the detection of B-mode polarization consistent with primordial gravitational waves with a tensor-to-scalar ratio, $r=0.2^{+0.07}_{-0.05}$, challenged predictions from most inflationary models of a lower value for $r$. More recent results by Planck on polarized dust emission show that the observed tensor modes signal is compatible with pure foreground emission. A more significant constraint on $r$  was then obtained by a joint analysis of Planck, BICEP2 and Keck Array data showing an upper limit to the tensor to scalar ratio $r\le 0.12$, excluding the case $r=0$ with low statistical significance. Forthcoming measurements by BICEP3, the Keck Array, and other CMB polarization experiments, open the possibility for making the fundamental measurement of $r$. Here we discuss how $r$ sets the scale for models where the dark matter is created at the inflationary epoch, the generically called super-heavy dark matter models. We also consider the constraints on such scenarios given by recent data from ultrahigh energy cosmic ray observatories which set the limit on super-heavy dark matter particles lifetime. We discuss how super-heavy dark matter can be discovered by a precise measurement of $r$ combined with future observations of ultra high energy cosmic rays.} 

\begin{document}
\maketitle 

\section{Introduction}
\label{sec:intro}

Most of the mass in the Universe is composed of Dark Matter (DM), matter observed only through its gravitational interaction and not through radiation. Both astrophysics and cosmology provide ample evidence for the existence of DM and its prevalence over visible matter (see e.g., \cite{Bergstrom:2000pn,Bertone:2004pz}). From the 1930's, astrophysicists have observed dynamics in large systems that can only be explained by dark matter such as  the dynamics of galaxy clusters \cite{Zwicky:1933gu}, galaxy rotation curves, and more recent multi-wavelength and gravitational lensing studies of interacting clusters, such as the ``bullet cluster" \cite{Clowe:2006eq}. On the cosmological side,  global fits to cosmological parameters enable a precise determination of the amount of DM present in the Universe. A recent result comes from Planck observations of the cosmic microwave background (CMB) which find the mean DM density in the Universe in units of the critical density to be $\Omega_{DM} h^2 = 0.1198 \pm 0.0015$ \cite{Planck:2015xua}, with $h = 0.678 \pm 0.009 ~{\rm km} {\rm s}^{-1} {\rm Mpc}^{-1}$, so $\Omega_{DM} = 0.261$. These indications, not only provide evidence for the presence of some unknown form of matter, but also indicate that this matter should be in the form of non-relativistic (cold) particles \cite{Planck:2015xua}. 

The leading paradigm to explain DM observations is based on the Weakly Interactive Massive Particle (WIMP) hypotheses, which consists of two basic assumptions: (i) WIMPs are stable particles of mass $M_\chi$ that interact weakly with the Standard Model (SM) particles; (ii) WIMPs are thermal relics, i.e. they were in Local Thermal Equilibrium (LTE) in the early Universe. Imposing that the WIMP density today is at the observed level, using a simple unitarity argument for the WIMP annihilation cross section $\sigma_{ann}\propto 1/M_{\chi}^2$, one obtains an estimate of the WIMP mass in the range of $10^{2}$ to  $10^{4}$ GeV. This result, also called the WIMP ``miracle", links the DM problem to the new physics scale expected in the context of the ``naturalness" argument for electroweak physics. Triggering in this way the strong hope that the search for WIMP DM may be connected to the discovery of new physics at the TeV scale. 

Searches for WIMP particles are ongoing through three different routes: direct detection, indirect detection, and accelerator searches (see e.g., \cite{Bertone:2004pz}). None of these efforts have discovered a clear WIMP candidate so far. In addition, no evidence for new physics has been observed at the Large Hadron Collider (LHC). Although not yet conclusive, the lack of evidence for WIMPs may imply a different solution for the DM problem outside of the WIMP paradigm. 

In this paper we will reconsider the scenario based on particle production due to time varying gravitational fields: the so-called Super Heavy Dark Matter (SHDM) scenario. This alternative approach is based on the possibility of particle production due to the non-adiabatic expansion of the background space-time acting on the vacuum quantum fluctuations. In quantum theories the possibility of producing particle pairs by the effect of a strong (classical) external field is well known: for instance, consider the case of $e^{\pm}$ pair creation by strong electromagnetic fields. The idea to apply such a mechanism in cosmology through external strong gravitational fields dates back to E. Schr\"odinger in 1939 \cite{Schro:1939}.

The construction of a coherent theory of particle creation by the expansion of the Universe was developed in the last 40 years and started with the pioneering works of \cite{Chernikov:1968zm,Parker:1968mv,Grib:1970mv,Zeldovich:1971mw,Grishchuk:1974ny}. More recently, in the framework of inflationary cosmologies, it was shown that particle creation is a common phenomenon, not tied to any specific cosmological scenario, that can play a crucial role in the solution to the DM problem as SHDM (labeled by $X$) can have $\Omega_X(t_0)\lesssim 1$ \cite{Chung:1998zb,Kuzmin:1998uv,Kolb:1998ki,Chung:1999ve,Chung:2001cb,Kolb:2007vd,Fedderke:2014ura}. This conclusion can be drawn under three general hypotheses: (i) SHDM in the early Universe never reaches LTE; (ii) SHDM particles have mass of the order of the inflaton mass, $M_{\phi}$; and (iii) SHDM particles are long-living particles with a lifetime exceeding the age of the Universe, $\tau_X\gg t_0$. 

Precision measurements of CMB polarization and observations of Ultra High Energy Cosmic Rays (UHECR) up to energies $\gtrsim 10^{20}$ eV  enable a direct experimental test of the three SHDM hypothesis listed above. The aim of the present paper is to discuss this issue in light of the latest observations.  

The observation of CMB fluctuations can be directly linked to the primordial density perturbations in the early Universe, which provide the initial condition for structure formation (see e.g. \cite{Lyth:1998xn,Bartolo:2004if}). Density perturbations are of two types: curvature and isocurvature,  the first one is connected with the total energy density in the early Universe while the second one to the actual composition of the energy density itself. The fact that SHDM particles with mass $M_X\sim M_\phi$ never attain LTE implies a non negligible contribution of the particles creation mechanism to isocurvature perturbations, that become source of gravitational potential contributing to the primordial gravitational wave background. 

Original reports from BICEP2 signalled the possibility of a substantial contribution of tensor modes to the CMB fluctuations \cite{Ade:2014xna} providing additional motivation for the SHDM production mechanism. As we discuss in section \ref{SHDM}, the level of tensor modes enables the determination of the energy scale and mass of the inflaton field resulting in a present time SHDM density at the required cosmological level. The result presented by BICEP2 appears to be seriously polluted by the polarised light emitted by dust at galactic scales \cite{Adam:2014bub}. The recent joint analysis by BICEP2, Keck Array and Planck \cite{Ade:2015tva}, validated on simulations of a dust-only modelling and performed by a simple subtraction of scaled spectra, have set the scale of $r$ at $r\le 0.12$ at $95\%$ confidence level. The final result of this analysis is expressed in terms of a likelihood curve for $r$ that peaks at $r=0.05$ but disfavours zero with a scarce statistical significance. As shown in \cite{Ade:2015tva}, this result can occur by chance $8\%$ of the time, as confirmed by dust-only simulations, and cannot be considered as a conclusive detection of primordial B-modes. In the near future several different detectors, both ground-based and sub-orbital, will take data at a variety of frequency enabling a more precise determination of the CMB primordial tensor modes. 

The third and last hypothesis of the SHDM paradigm is a long particle life-time, much longer than the age of the Universe. This is a general requirement of any DM model based on the existence of new particles, common to the WIMP hypothesis as well. As in the case of WIMPs, discrete gauge symmetries protecting particles from fast decays work equally well for SHDM particles. A general particle physics example of SHDM with life-time exceeding the age of the Universe can be found in \cite{Ellis:1990iu}. 

The best way to test the existence of SHDM is through the indirect detection of its decay and/or annihilation products since direct detection of SHDM is unattainable. For instance, the annihilation of SHDM can occur in the case they are "majorana-like" particles, i.e. particles and anti-particles coincide. Since the annihilation cross section of a (point) particle is bounded by unitarity, $\sigma_{ann}\propto 1/M_X^2\sim 1/M_\phi^2$, the annihilation process results in an unobservably small annihilation rate \cite{Aloisio:2006yi}. Even if alternative theoretical models can be constructed, with a sizable annihilation rate \cite{Blasi:2001hr}, we will not consider this case here and focus on the more general case of signatures from SHDM decays.

If SHDM particles decay, we can determine the composition and spectra of the SM particles produced by SHDM decays under general assumptions on the underlying theory \cite{Aloisio:2003xj,Aloisio:2006yi}. Given the mass $M_X\sim M_\phi$ and the lifetime $\tau_X\gg t_0$,  UHECR experiments with high statistics should be able to detect the decay products at extreme energies $\gtrsim 10^{20}$ eV. In section \ref{UHECR} we discuss the possibility of indirect detection of SHDM particles focusing on spectra and chemical composition taking into account the latest UHECR observations performed by Auger \cite{ThePierreAuger:2013eja,Aab:2014aea,Aab:2014kda} and Telescope Array (TA) \cite{Abu-Zayyad:2013qwa}. 

When CMB polarization experiments determine the inflation scale, UHECR detectors will study the parameter space for SHDM lifetimes. In section \ref{Conclu} we will draw our conclusions discussing the future possibilities in constraining SHDM models through CMB and UHECR observations, highlighting the discovery capabilities of the next generation UHECR experiments. 

\section{Tensor modes and Super Heavy Dark Matter} 
\label{SHDM}

In the scenario based on gravitational production, SHDM never attains LTE and we can connect its density today with the density when it was created. Following \cite{Chung:1998zb} we can write 

\begin{equation}
\frac{\rho_X(t_0)}{\rho_R(t_0)}=\frac{\rho_X(t_{RH})}{\rho_R(t_{RH})} \left (\frac{T_{RH}}{T_0} \right )
\label{eq:den1}
\end{equation}
where $\rho_R$ is the energy density of radiation, $\rho_X$ is the SHDM energy density, and $T_{RH}$ and $T_0$ are the temperature of the Universe at reheating time $t_{RH}$ and today $t_0$, respectively. Assuming that SHDM particles are produced just after the de Sitter phase at time $t_e$, when the coherent oscillation phase begins, then the inflaton energy density and the SHDM energy density will be redshifted at approximately the same rate. This holds until the reheating phase finishes and the radiation dominated phase begins. Therefore, assuming that before reheating most of the energy density of the Universe is converted into radiation \cite{Chung:1998zb} ($\rho_R\simeq \rho_c=3H^2M_{Pl}^2/8\pi$), we can rewrite 
\begin{equation}
\frac{\rho_X(t_{RH})}{\rho_R(t_{RH})} \simeq \frac{8\pi}{3}\frac{\rho_X(t_e)}{M_{Pl}^2H^2(t_e)}
\label{eq:den2}
\end{equation}
where $M_{Pl}=10^{19}$ GeV (the Planck mass), and $H(t_e)$ is the Hubble parameter at $t_e$.

The energy density of particles produced by the interaction of a (time varying) classical gravitational field with a quantum vacuum depends on the details of the underlying quantum theory (e.g., the choice of vacuum) and cosmology (e.g., the structure of the time dependence of the scale factor). These details were analyzed by several authors in the framework of different models, however, as was shown in \cite{Chung:1998zb,Kuzmin:1998uv,Kuzmin:1999zk}, the SHDM density at the end of inflation can be generally approximated by  \cite{Kuzmin:1999zk,Chung:2004nh}
\begin{equation}
\rho_{X}(t_e)\sim 10^{-3} M_X^4 \left (\frac{M_X}{H(t_e)} \right )^{-3/2} e^{-2M_X/H(t_e)}~,
\label{eq:rhoX}
\end{equation}
quite independent of the specific underlying theory. Eq. (\ref{eq:rhoX}) is inaccurate in the low mass limit $M_X \ll M_\phi$, while it works reasonably well near the cut-off region $M_X \gtrsim M_\phi$ that is the regime that we are interested in (see later). For a more accurate determination of the SHDM density, obtained by numerically solving mode equations, taking into account also the isocurvature-curvature cross correlation, see \cite{Chung:2011xd,Chung:2013sla}.

Using Eq. (\ref{eq:den1}), (\ref{eq:den2}) (\ref{eq:rhoX}), and  taking the Hubble parameter at $t_e$ equal to the inflaton mass $H(t_e)\sim M_\phi$ one gets a simple estimate of the ratio of the SHDM density to the critical density today $\Omega_X=\rho_X(t_0)/\rho_c(t_0)$: 
\begin{equation}
\Omega_X(t_0)\simeq 10^{-3} \Omega_R \frac{8\pi}{3} \left(\frac{T_{RH}}{T_0} \right ) \left (\frac{M_\phi}{M_{Pl}} \right)^2 \left (\frac{M_X}{M_\phi}\right)^{5/2}  e^{-2M_X/M_{\phi}}~.
\label{eq:omX}
\end{equation}
being $T_0=2.3\times 10^{-13}$ GeV the CMB temperature today and $\Omega_R=4\times 10^{-5}$ the radiation density today. 

The gravitational production of SHDM is intimately connected to the process that, during inflation, generated primordial large-scale density fluctuations. As realized in \cite{Chung:2004nh}, the observations of the CMB fluctuations provide important insights into SHDM properties having the potential to falsify the model. The production of SHDM during inflation gives rise to isocurvature perturbations that become sources of gravitational potential energy contributing to the tensor power spectrum of the CMB. 

The observation of a sizeable amount of tensor modes in the CMB fluctuation pattern would imply an important confirmation of the inflationary paradigm in agreement with the hypothesis that adiabatic perturbations originate within the single field, slow-roll framework of inflation. 

To asses the impact of primordial B-modes on SHDM, we express the inflaton mass $M_{\phi}$ in terms of the tensor to scalar ratio $r$. At leading order in the slow roll approximation we can write the inflaton potential height $V_{\star}$ in terms of $r$ and the amplitude $A_s$ of the scalar perturbations on super-Hubble scales. Using the combined analysis of Planck and WMAP data one has \cite{Ade:2015lrj}:
\begin{equation}
V_\star \simeq \frac{3\pi^2}{2} A_s r M_{Pl}^4 \simeq (2\times 10^{16} {\rm GeV})^4 \frac{r}{r_0}=M_{GUT}^4  \left (\frac{r}{r_0} \right )
\label{eq:Vstar}
\end{equation}
being $M_{GUT}=2\times 10^{16}$ GeV the scale of Grand Unification Theory (GUT) and $r_0=0.12$ is the experimental upper bound on the tensor to scalar ratio \cite{Ade:2015lrj}. It is interesting to note here that in the case of a substantial contribution of B-modes, as in the case of the BICEP2 claim \cite{Ade:2014xna}, one finds an inflaton potential height intriguingly near the typical GUT scale. Let us now assume an inflaton potential of the type: 
\begin{equation}
V(\phi)=\frac{M_{\phi}^{4-\beta}}{\beta} \phi^{\beta}
\label{eq:Vphi}
\end{equation}
where $\beta=2/3,1,4/3,2$. From the latest Planck observations \cite{Ade:2015lrj} follows that the quadratic potential is moderately disfavoured while higher values of $\beta$ are strongly disfavoured. Values of $\beta=2/3,1,4/3$ are motivated by axion monodromy, which combines chaotic inflation and natural inflation \cite{Linde:1983gd,Freese:1990rb,Silverstein:2008sg,McAllister:2014mpa}.

\begin{figure}
   \centering
   \includegraphics[width=0.495\textwidth]{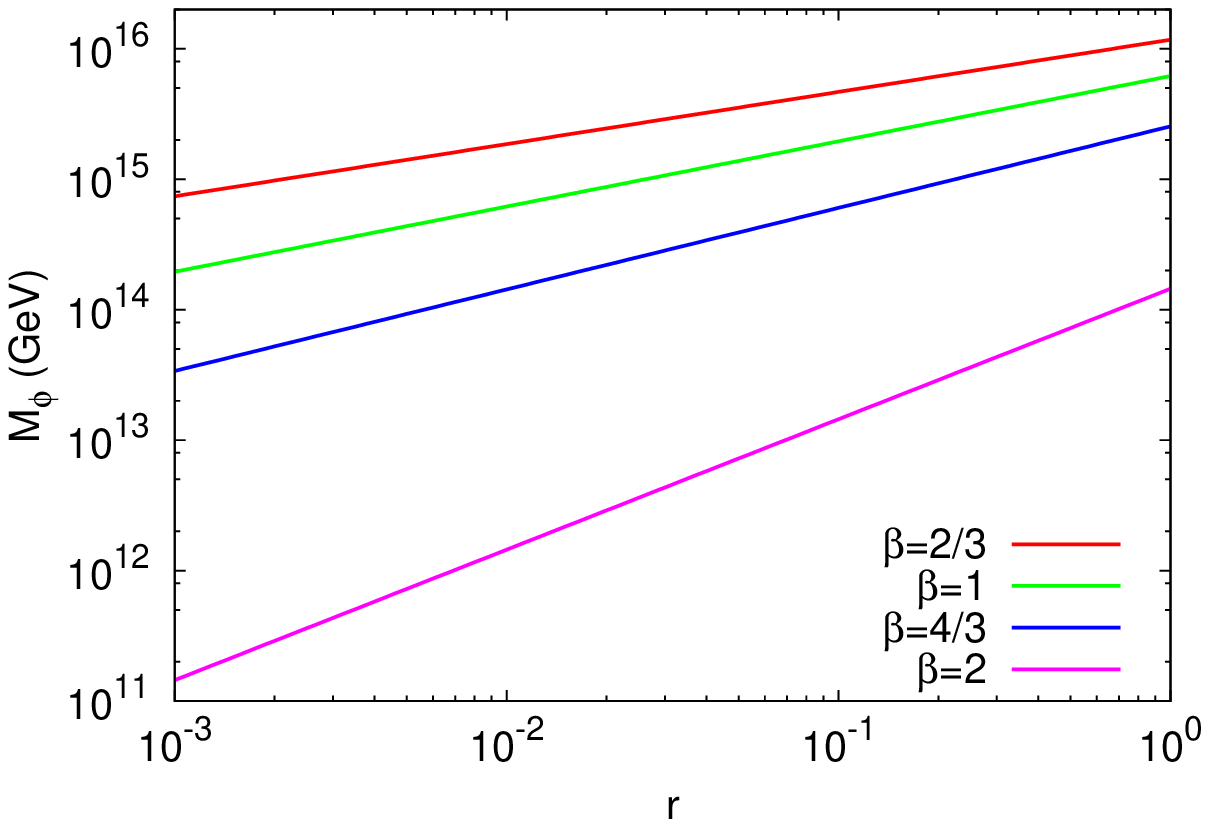}
   \includegraphics[width=0.495\textwidth]{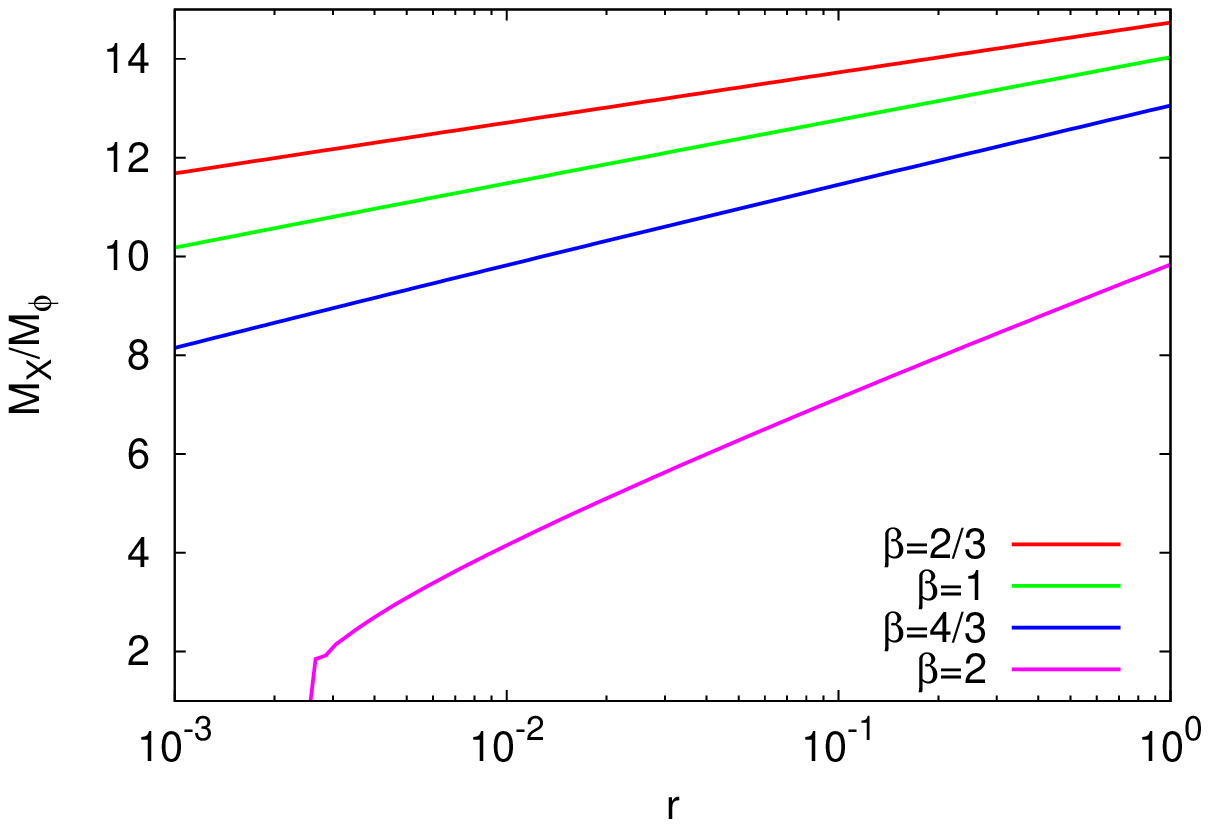}
   \caption{ [Left panel] Inflaton mass as function of the ratio $r$ of tensor to scalar modes for different choices of the inflaton potential as labeled. [Right panel] Ratio of the SHDM mass and inflaton mass as function of $r$, obtained as solution of the equation $\Omega_X=\Omega_{DM}$ using different choices of the inflaton potential as labelled.}
   \label{fig0} 
\end{figure}

From Eqs. (\ref{eq:Vstar}) and (\ref{eq:Vphi}), using the definition of the slow-roll parameter $\epsilon(\phi)$
\begin{equation}
\epsilon(\phi) = \frac{M_{Pl}^2}{16\pi} \left [ \frac{V'(\phi)}{V(\phi)} \right ]^2 = \frac{r}{16}~,
\label{epsilon}
\end{equation}
we can determine the inflaton mass as a function of the tensor to scalar ratio $r$: 
\begin{equation}
M_{\phi} = M_{GUT} \left [ \beta \left (\frac{M_{GUT}}{M_{Pl}} \right)^\beta \left ( \frac{\sqrt{\pi r_0}}{\beta}\right )^\beta \left (\frac{r}{r_0}\right )^{1+\beta/2}\right ]^{\frac{1}{4-\beta}}~,
\label{eq:Mphi}
\end{equation}
In figure \ref{fig0}, left panel, we plot the value of the inflaton mass $M_{\phi}$ as function of r, choosing different values of the inflaton potential power law index $\beta=2/3,1,4/3,2$ \cite{Ade:2015lrj,Linde:1983gd,Freese:1990rb,Silverstein:2008sg,McAllister:2014mpa}.   

Using Eq. (\ref{eq:Mphi}) we can rewrite the density of SHDM today in terms of the inflaton potential power law index $\beta$, the ratio of tensor to scalar modes $r$ and the ratio $M_X/M_\phi$: 
$$
\Omega_X(t_0)\simeq 10^{-3}\Omega_R \frac{8\pi}{3} \left (\frac{T_{RH}}{T_0} \right ) \times ~~~~~
$$
\begin{equation}
\times \beta^{\frac{2}{4-\beta}} \left (\frac{M_{GUT}}{M_{Pl}} \right )^{\frac{8}{4-\beta}} \left (\frac{\sqrt{4\pi r_0}}{\beta} \right )^\frac{2\beta}{4-\beta} \left (\frac{r}{r_0} \right )^\frac{2+\beta}{4-\beta} \left (\frac{M_X}{M_\phi}\right )^{5/2} e^{-2M_X/M_\phi}~.
\label{eq:omXfinal}
\end{equation}

Taking the reheating temperature $T_{RH}\simeq 10^{9}$ GeV and assuming that the SHDM density today coincides with the observed DM density $\Omega_X=\Omega_{DM}=0.261$, using Eq. (\ref{eq:omXfinal}), we can determine the ratio $M_X/M_\phi$ as function of $r$ for different choices of the inflaton potential. In the right panel of figure \ref{fig0} we plot the resulting behaviours. As already discussed in \cite{Chung:2004nh}, although only in the case of an inflaton potential with $\beta=2$, the typical values of the SHDM mass are such that $M_X/M_\phi \gtrsim O(5)$ with a range in mass that spans from $10^{12}$ GeV up to $10^{17}$~GeV, depending on the value of $r$ and the choice of the inflaton potential power law index $\beta$. 

The result presented in the right panel of figure \ref{fig0} also depends on the assumption about the reheating temperature. For instance, in the case $\beta=2$ the equation $\Omega_X=\Omega_{DM}=0.261$ has no solutions for $r<3\times 10^{-3}$, in this case a larger value\footnote{In inflationary scenarios embedded in gravity mediated SUSY breaking the gravitino over-production bound restricts the reheating temperature to $T_{RH}\lesssim \times10^{9}\div 10^{10}$ GeV that in the case $\beta=2$ corresponds to a lower bound on the tensor to scalar ratio around $10^{-3}$ (see \cite{Chung:2004nh} and references therein).}  of $T_{RH}$, at the level of $T_{RH}\simeq 10^{10}$ GeV, is needed in order to obtain the correct SHDM density today for $r\lesssim 3\times 10^{-3}$. 

To assess the dependence on the reheating temperature of the result presented in the right panel of figure \ref{fig0}, we have repeated that computation using different values of $T_{RH}$. In the case $\beta=2$, as discussed in \cite{Chung:2004nh}, there is a lower bound on $T_{RH}$, fixing $T_{RH}$ at its minimum the equation $\Omega_X=\Omega_{DM}$ shows no solution only for a tensor to scalar ratio $r<2\times 10^{-2}$, while at larger $r$ the ratio $M_X/M_\phi$ differs by no more than $50\%$ from the result of figure \ref{fig0}. 
In the case $\beta=2/3,1,4/3$ there is no lower bound on the $T_{RH}$ value. In these cases, to check the dependence of our results on $T_{RH}$ we have repeated the computations of figure \ref{fig0} (right panel) taking different values of $T_{RH}$ down to $10^{5}$ GeV always obtaining solutions $M_X/M_\phi$ that differ by no more than $50\%$ from the values of figure \ref{fig0} (where $T_{RH}=10^9$ GeV). A change in the $M_X$ parameter by a factor of two, changing the $T_{RH}$ value, does not affect the result presented here, therefore in the following we will consider only the case $T_{RH}=10^{9}$ GeV as a reference value. 

Another important point concerns the limits on the the SHDM models that come from the experimental limits on isocurvature perturbations. Almost independently of $T_{RH}$, we found that the ratio $M_X/M_\phi$ needed to obtain the observed DM density is in the range $2\div15$. With this parameters choice, as shown by \cite{Chung:2004nh} and more recently by \cite{Chung:2011xd}, the isocurvature perturbations expected are well within the experimental limits.

We can conclude this section by stating that super-heavy particles produced by time varying gravitational fields in the early Universe provide a viable explanation of the DM problem even with a ratio of tensor to scalar modes at the level of $10^{-3}$. This is a non trivial result that shows how the observation of a non zero fraction of tensor modes in the CMB fluctuations pattern, already at the level of $10^{-3}\div 10^{-2}$, confirms that the gravitational production of SHDM particles is a viable mechanism to explain the DM problem, assuring a density of SHDM today at the observed level. In the next section we will consider the case of SHDM discussing its possible indirect detection through UHECR observations. 

\section{Ultra High Energy particles by Super Heavy Dark Matter decay}
\label{UHECR}

In order to constitute a substantial fraction of the observed DM, SHDM should be composed by metastable particles with lifetime longer or much longer than the age of the Universe. In the previous section we have discussed the gravitational production of SHDM showing how the mass of these particles is fixed by cosmology. We are left with a single free parameter in the model: the particle lifetime $\tau_X$, that can be constrained through observations of the decay products. 

Following \cite{Aloisio:2003xj,Aloisio:2006yi}, on very general grounds, we can assume that the SHDM decay produces a pair of quark anti-quark that gives rise to a parton cascade, producing SM particles through the subsequent hadronization process. The basic signatures of the SHDM decay process are three: (i) SHDM particles (as any other DM particle) cluster gravitationally and accumulate in the halo of our Galaxy with an average density of $\rho_X^{halo}\simeq 0.3$ GeV/cm$^3$; (ii) in the hadronic cascades the most abundant particle produced are pions, therefore UHE neutrinos and photons are the most abundant particles expected on Earth; (iii) the non-central position of the sun in the Galactic halo results in an anisotropic flux of the  SHDM decay products  \cite{Dubovsky:1998pu,Aloisio:2007bh}.

The quantitative predictions for energy spectra and chemical composition of UHECR coming from SHDM decay require an extrapolation of QCD calculations from the TeV scale up to the scale $M_X$. There are several computational schemes proposed in the literature based on analytic approximations \cite{Berezinsky:1998ed} or Monte Carlo simulations \cite{Birkel:1998nx,Berezinsky:2000up,Sarkar:2001se,Barbot:2002gt,Aloisio:2003xj}. These schemes predict accurately the secondary spectra of SM particles produced in the SHDM decay and agree well each other. Their most important outcome from the observational point of view is a flat spectrum, that at the relevant energies can be approximated as $dN/dE\propto E^{-1.9}$, independently of the particle type, with a photon/nucleon ratio of about $\gamma/N \simeq 2\div 3$ and a neutrino/nucleon ratio $\nu/N\simeq 3\div 4$, quite independent of the energy \cite{Aloisio:2003xj}.

The UHECR emissivity produced by the decay of SHDM in the halo of our Galaxy can be written as 

\begin{equation}
I_{p,\gamma,\nu}(E)=\frac{1}{M_X\tau_X}\frac{dN_{p,\gamma,\nu}(E)}{dE}\rho_X(R)
\label{emiss}
\end{equation}
where $M_X$ and $\tau_X$ are the mass and lifetime of the SHDM particle, $dN/dE$ is the energy spectrum (fragmentation function) of the SM particles produced by the decay and $\rho_X(R)$ is the density of X-particles in the Galaxy as function of the distance $R$ from the Galactic Center (GC).

In the present paper, assuming that SHDM contributes to a substantial fraction of the galactic DM, we will use the DM density profile as computed in several numerical simulations by Navarro-Frenk-White (NFW) \cite{Navarro:1996gj} and Moore \cite{Moore:1999nt}: 

\begin{equation} 
\rho_X(R)=\frac{\rho^0_X}{(R/R_s)^\alpha (1+R/R_s)^{3-\alpha}}
\label{DMdensity}
\end{equation} 
with $\alpha=1,3/2$ in the case of NFW and Moore respectively and $R_s=45$ kpc being the fiducial radius of the DM distribution as obtained in \cite{Berezinsky:1998rp}. The DM density is normalized to the expected local (solar) density, namely: $\rho_X(R_\odot) = \rho^{DM}_\odot = 0.3$ GeV/$cm^3$. 

The expected flux on Earth is given by the integral over the line of sight of the emissivity (\ref{emiss}), indicating with $\theta$ the angle between the line of sight and the axis Earth-GC we can write the flux as 

\begin{equation} 
J_i(E,\theta)=\frac{1}{4\pi}\int_0^{s_{max}(\theta)} ds I_i(E,s)
\label{flux1}
\end{equation} 
with 
$$s_{max}(\theta)=R_\odot \cos(\theta)+ \sqrt{R_H^2 + R_{\odot}^2 sin^2\theta},$$
being $i=p,\gamma,\nu$ the type of particle produced in the SHDM decay, $s$ the coordinate along the line of sight, $R_\odot = 8.5$  kpc the distance between Earth and GC and $R_H = 100$ kpc the radius of the galactic halo. Changing integration variable in equation (\ref{flux1}) from the line of sight coordinate $s$ to the galactic radius $R$, one obtains a more convenient formula for numerical computations \cite{Aloisio:2006yi}: 

\begin{equation}
J_i(E,\theta)=\frac{1}{4\pi\tau_X M_X} \frac{dN_i(E)}{dE} \left [ 
2\int_{R_\odot sin(\theta)}^{R_\odot} dR R\frac{\rho_X(R)}{\sqrt{R^2 - R^2_\odot sin^2(\theta)}} + 
\right .
\end{equation}
$$ \left . + \int_{R_\odot}^{R_H} dR R\frac{\rho_X(R)}{\sqrt{R^2 - R^2_\odot sin^2(\theta)}} \right ].$$

Together with UHECR emitted by the decay of SHDM in our local halo (\ref{flux1}) one should take into account also the contribution of the overall Universe to the emitted flux of particles. In the case of protons and photons the energy losses due to the interactions with astrophysical backgrounds (mainly CMB) reduce the expected cosmological flux to a negligible fraction with respect to the flux produced by our local halo \cite{Aloisio:2006yi}. In the case of neutrinos the contribution to the expected flux from the whole Universe is not negligible being at a level of about $10 \%$ of the local flux. Therefore one should add to the neutrino flux given by (\ref{flux1}) also the flux given by:
\begin{equation} 
J^{EG}_{\nu}(E)=\frac{c}{4\pi}\frac{\Omega_{X}\rho_c}{M_X\tau_X}\int_0 dz \left | \frac{dt}{dz} \right | \frac{dN_{\nu}}{dE}[(1+z)E]e^{-S_{\nu}(E,z)}
\label{flux2}
\end{equation}  
where $dt/dz=-H_0\sqrt{(1+z)^3\Omega_m + \Omega_{\Lambda}}$, $\rho_c=5.5\times 10^{-6}$ GeV/cm$^3$ is the critical density of the Universe and $S_\nu(E_\nu,z)$ is the neutrino opacity of the Universe calculated in \cite{Gondolo:1991rn,Weiler:1982qy}. 

\begin{figure}
   \centering
   \includegraphics[width=0.495\textwidth]{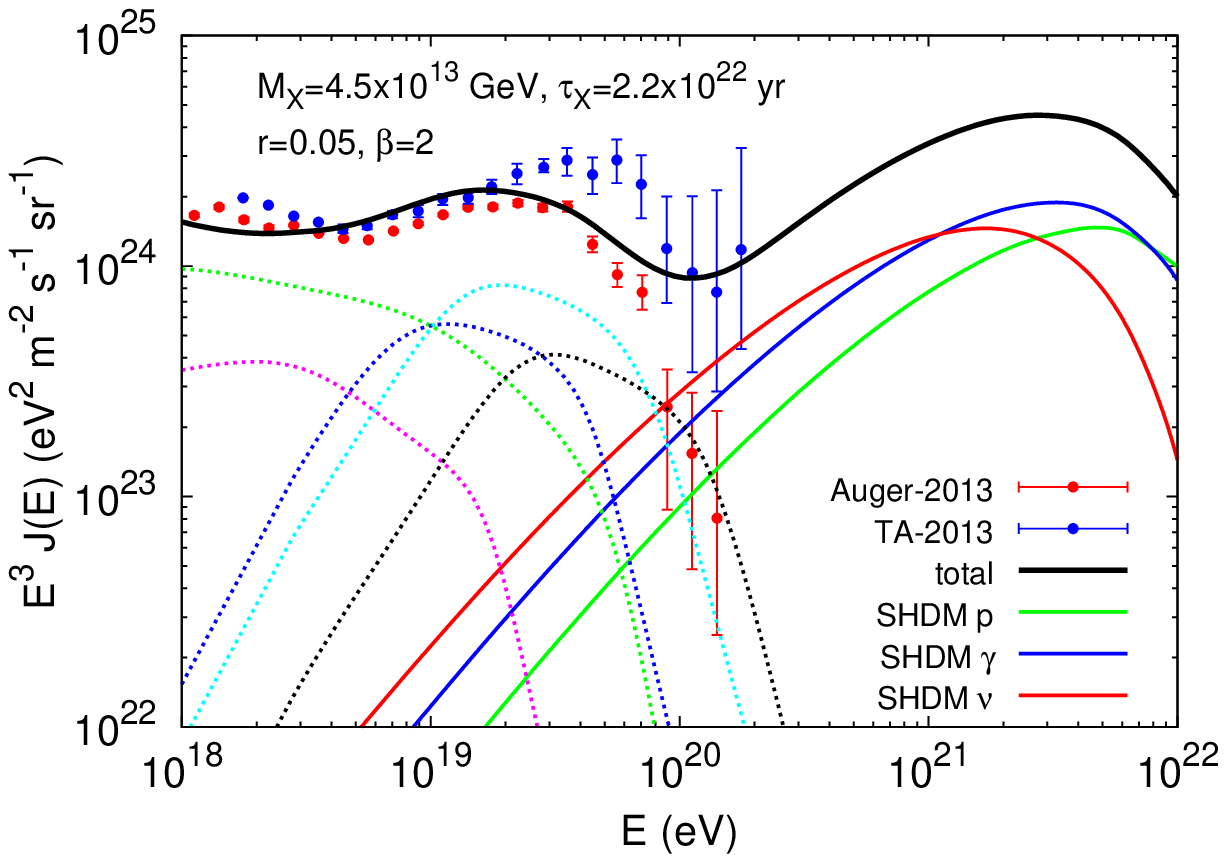}
   \includegraphics[width=0.495\textwidth]{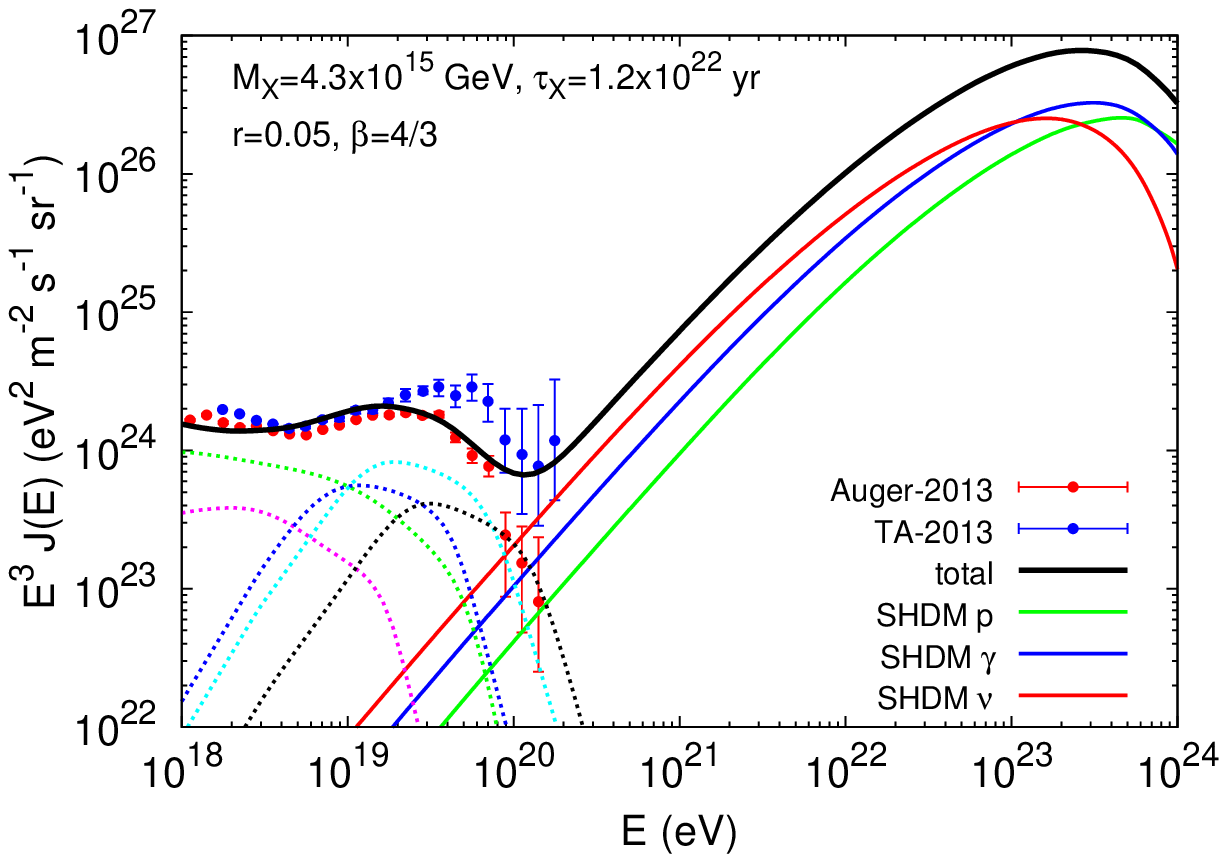}
   \includegraphics[width=0.495\textwidth]{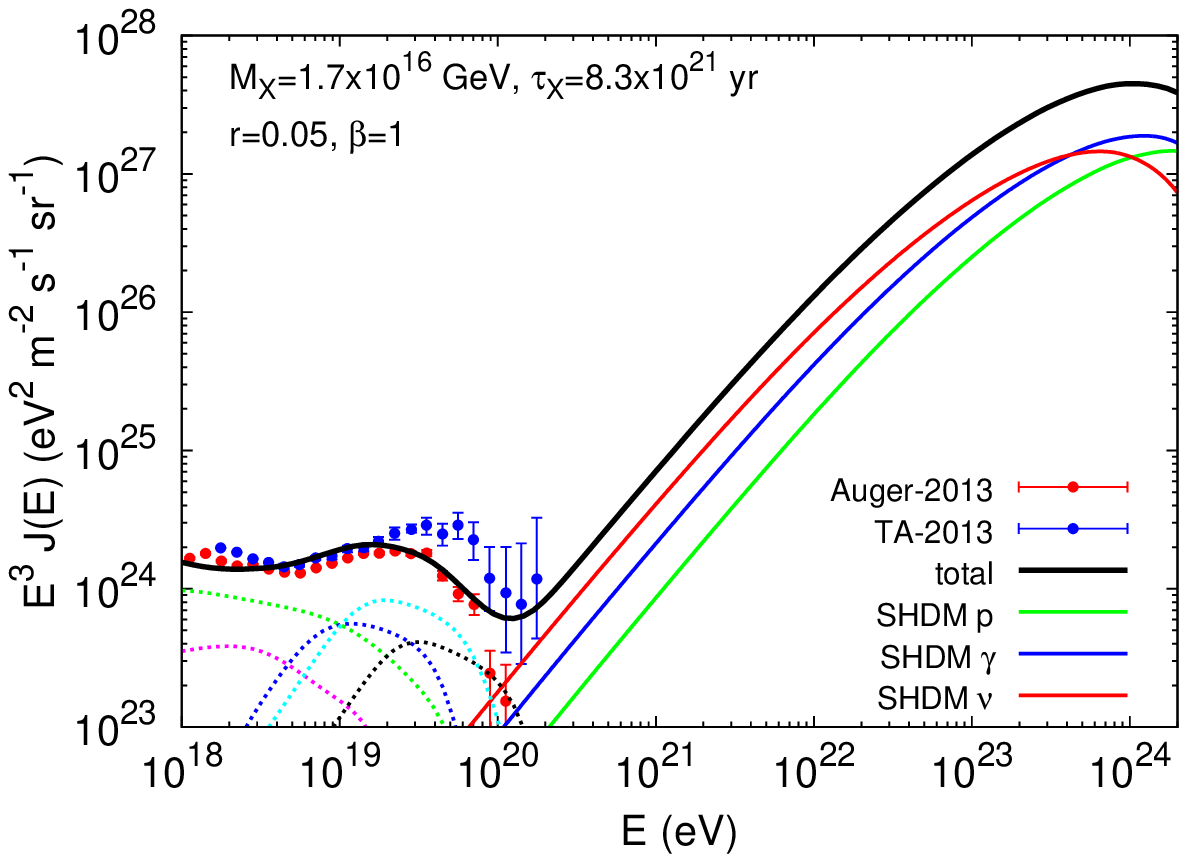}
   \includegraphics[width=0.495\textwidth]{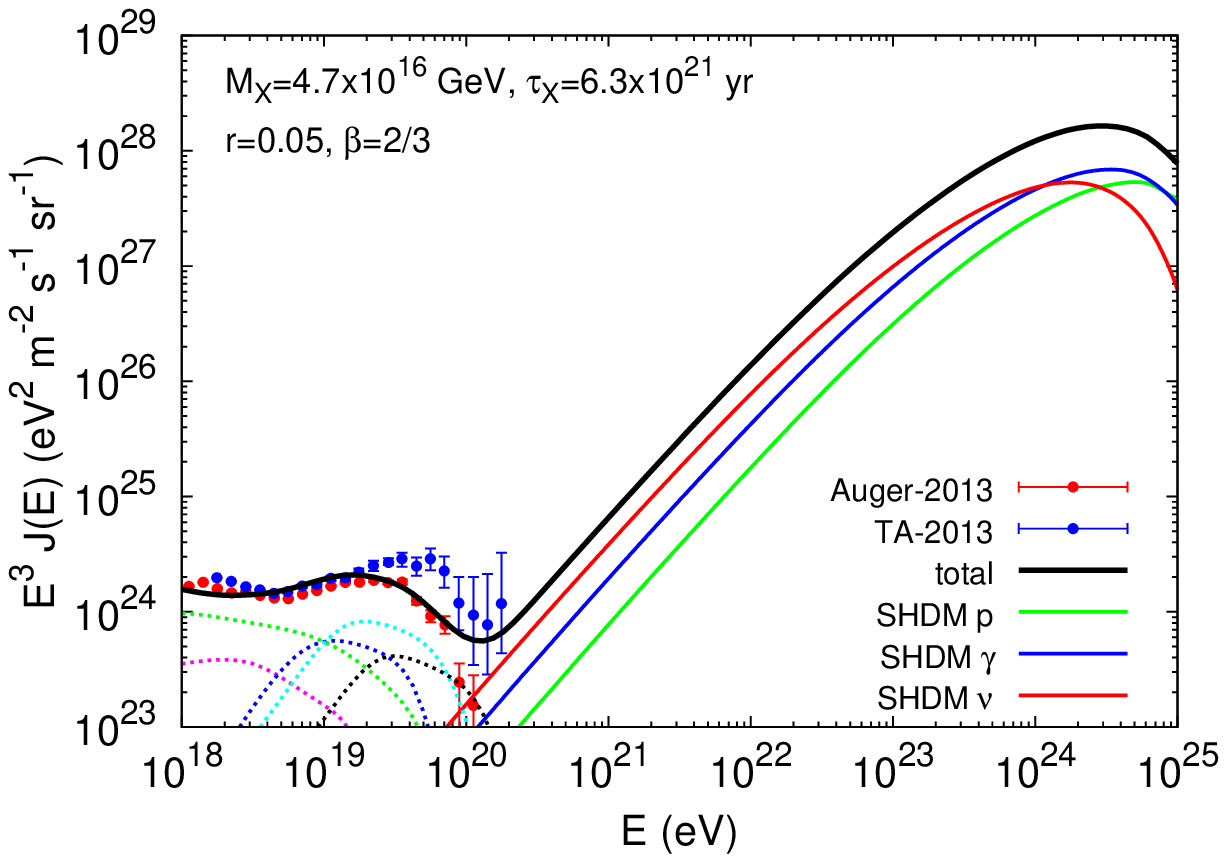}
   \caption{ UHECR flux: proton (dotted green), He (dotted magenta), CNO (dotted blue), MgAlSi (dotted cyan), Fe (dotted black) from astrophysical sources \cite{Aloisio:2013hya} and proton (green solid), photon (blue solid) and neutrino (red solid) from the decay of SHDM with a Moore density profile. The total flux is represented by the black thick solid line. Experimental data are the latest observations of Auger \cite{ThePierreAuger:2013eja} and TA \cite{Abu-Zayyad:2013qwa}. All plots are obtained assuming $r=0.05$, taking the four different choices of the inflaton potential: $\beta=2$ upper left panel, $\beta=4/3$ upper right panel, $\beta=1$ lower left panel and $\beta=2/3$ lower right panel. The corresponding values of the SHDM parameters $(M_X,\tau_X)$ are labelled in the different panels.} 
   \label{fig1} 
\end{figure}

The contribution of SHDM decay to the UHECR fluxes starts to be relevant at the highest energies ($E>5\times 10^{19}$ eV). Using the model proposed in \cite{Aloisio:2013hya} for UHECR by astrophysical sources (see also \cite{Aloisio:2015ega,Aloisio:2012wj,Allard:2005ha,Allard:2008gj,Unger:2015laa,Globus:2015xga} for different computation schemes and models of UHECR composition), in figure \ref{fig1} we have plotted the total UHECR flux, highlighting the contribution of the different components: proton (dotted green), He (dotted magenta), CNO (dotted blue), MgAlSi (dotted cyan), Fe (dotted black) from astrophysical sources \cite{Aloisio:2013hya} and proton (green solid), photon (blue solid) and neutrino (red solid) from the decay of SHDM. The latter fluxes where normalized integrating over the whole sky ($0 \leqslant \theta < \pi$), with a Moore density profile for SHDM. The four panels of figure \ref{fig1} correspond to the four different assumptions on the inflaton potential discussed in the previous section $\beta=2/3,1,4/3,2$ (as labelled in the figure) and fixing a reference value of the tensor to scalar ratio $r=0.05$, that corresponds to the peak in the tensor to scalar ratio likelihood curve of the combined analysis of Planck, BICEP2 and Keck Array \cite{Ade:2015tva}. As labelled in the figure, this choice of the inflation parameters corresponds to a SHDM mass: $M_X=4.5\times 10^{13}$ GeV in the case of $\beta=2$, $M_X=4.3\times 10^{15}$ GeV with $\beta=4/3$, $M_X=1.7\times 10^{16}$ GeV with $\beta=1$ and $M_X=4.7\times 10^{16}$ GeV with $\beta=2/3$. The minimum lifetime of SHDM was fixed by imposing the Auger experimental limits on the $\gamma$ ray flux at $E\ge 10^{19}$ eV, observed at the level of $2\%$ \cite{Abreu:2011pf,Aglietta:2007yx}, and on the number of events at $E\ge 6\times 10^{19}$ eV \cite{ThePierreAuger:2013eja}. These minimum lifetime values are: $\tau_X=2.2\times 10^{22}$ yr in the case of $\beta=2$, $\tau_X=1.2\times 10^{22}$ yr in the case of $\beta=4/3$, $\beta=8.3\times 10^{21}$ yr in the case of $\beta=1$ and $\tau_X=6.3\times 10^{21}$ yr in the case of $\beta=2/3$.

Let us now take into account the composition of UHECR. The chemical composition is inferred by the mean value of the depth of shower maximum $\langle X_{max} \rangle$ and its dispersion (RMS) $\sigma (X_{max})$. As was first discussed in \cite{Aloisio:2007rc} (see also \cite{Kampert:2012mx}), the combined analysis of $\langle X_{max} \rangle$ and $\sigma(X_{max})$ is more sensitive to the chemical composition and provides less model dependent results. The main uncertainties in such a procedure are provided by the dependence of $\langle X_{max} \rangle$ and its fluctuations on the interaction models used for the shower development. Most of such models fit low energy accelerator data while providing somewhat different results when extrapolated to the energies of relevance for UHECRs (for a review see \cite{Engel:2011zzb} and references therein).

In our calculations we follow the procedure of \cite{Abreu:2013env} where four models of HE interaction were included to describe the atmospheric shower development, namely EPOS 1.99 \cite{Pierog:2006qv}, Sibyll 2.1 \cite{Ahn:2009wx}, QGSJet 01 \cite{Kalmykov:1997te} and QGSJet 02 \cite{Ostapchenko:2005nj}, in order to derive for each given nuclear primary a simple prescription for $\langle X_{max} \rangle $ and $\sigma(X_{max})$. To account also for the contribution of UHE photons to $X_{max}$ and $\sigma(X_{max})$ we have used CONEX simulations of $\gamma$-induced showers, taking into account the Landau Pomeranchuk Migdal (LPM) and the Geomagnetic field effects on the showers development \cite{Engel:2007zzc}.

The observations of Auger point toward a mixed composition of UHECR with a prevalent light composition at low energies $E<5\times 10^{18}$ eV and a progressively heavier composition at the highest energies \cite{ThePierreAuger:2013eja,Aab:2014aea,Aab:2014kda}. The highest energy bin with an observed chemical composition is placed at $E\simeq 5\times 10^{19}$ eV, at highest energies there are no available measurements of composition \cite{ThePierreAuger:2013eja,Aab:2014aea,Aab:2014kda}. As discussed above, the decay of SHDM gives a substantial contribution to the composition of UHECR only at energies larger than $5\times 10^{19}$ eV, where the fraction of photons and neutrinos rises over  a few $\%$, therefore in an energy band not scrutinized by Auger and TA because of the extremely low statistics of the collected events.  

\begin{figure}
   \centering
   \includegraphics[width=0.495\textwidth]{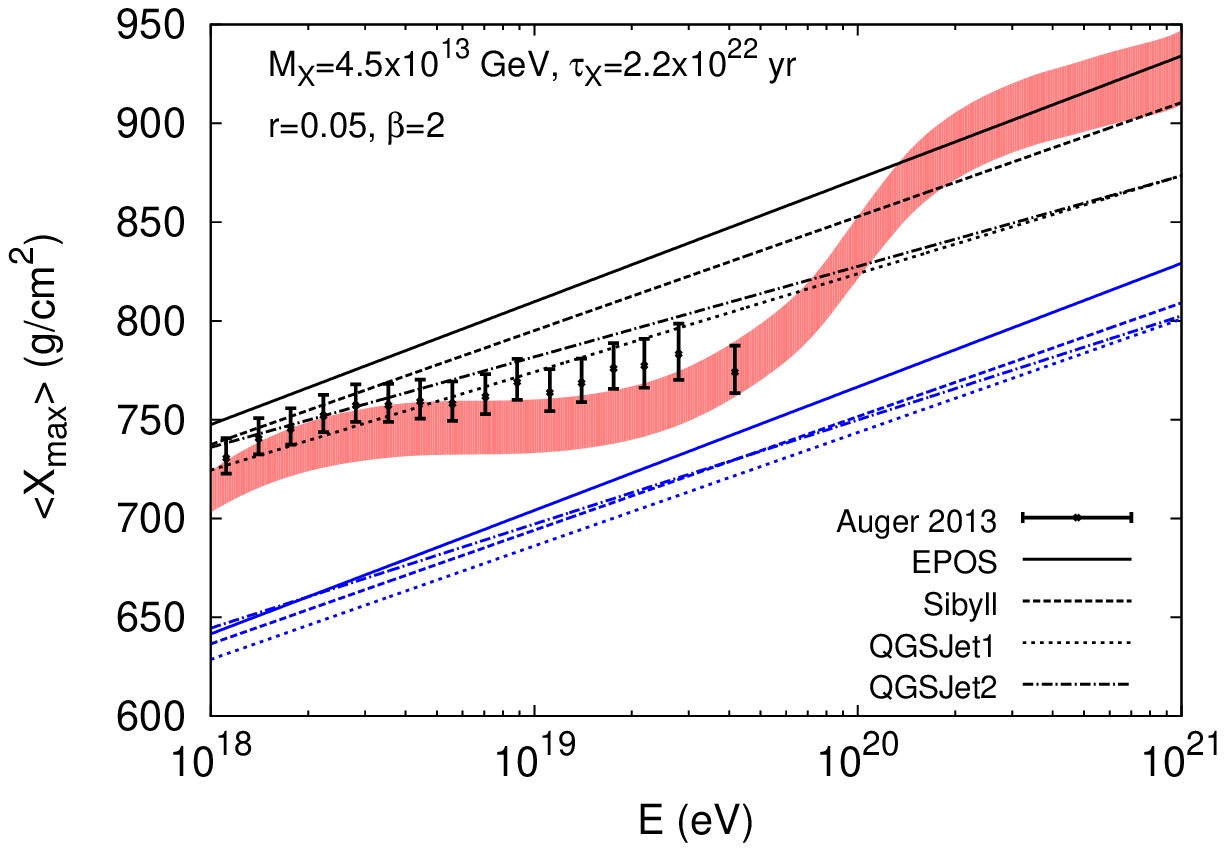}
   \includegraphics[width=0.495\textwidth]{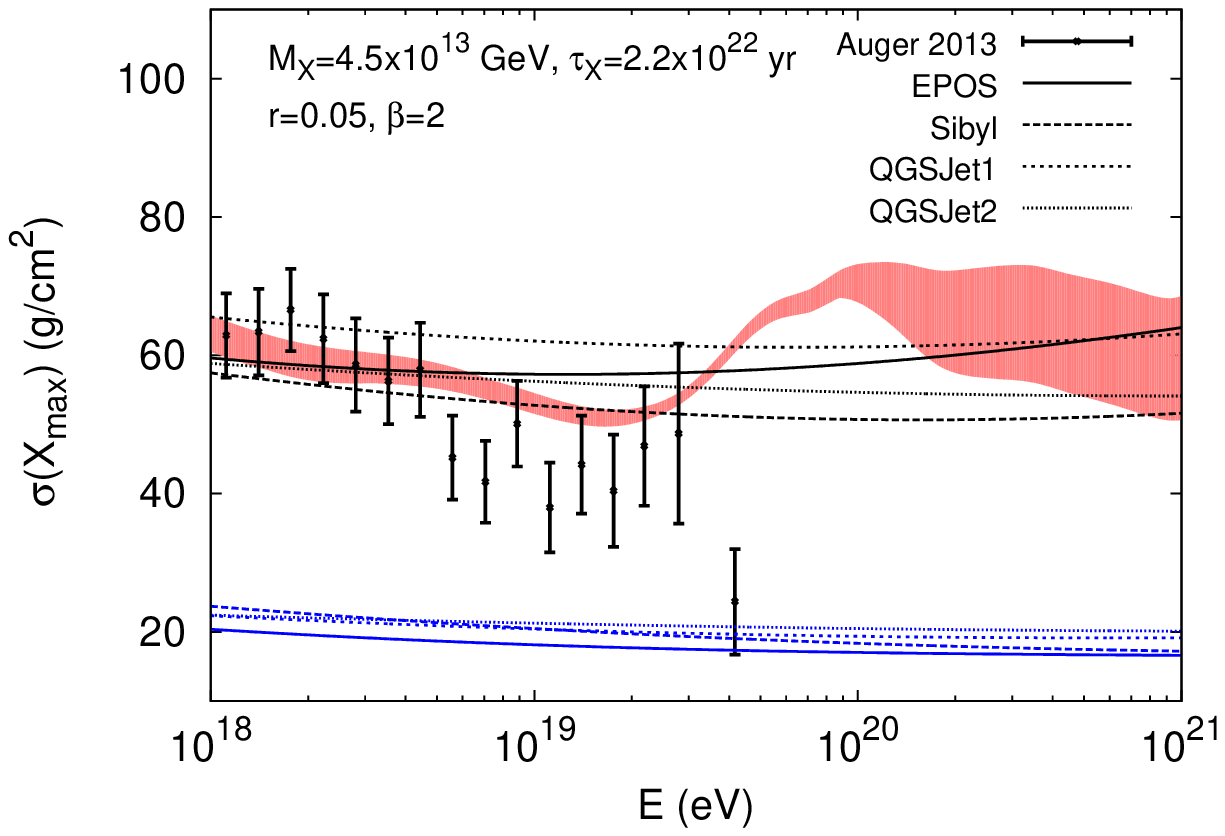}
   \caption{$\langle X_{max} \rangle $ (left panel) and $\sigma(X_{max})$ (right panel) as computed in this paper compared with the latest Auger observations \cite{ThePierreAuger:2013eja,Aab:2014aea,Aab:2014kda}. Shown by the red band is the uncertainty in the computations due to the choice of the hadron interaction model (see text).}
   \label{fig2} 
\end{figure}

In figure \ref{fig2} we plot $\langle X_{max} \rangle $ and $\sigma(X_{max})$ corresponding to the fluxes of the upper left panel of figure \ref{fig1}, obtained with a tensor to scalar ratio $r=0.05$ and an inflaton potential power law index $\beta=2$, together with the Auger data of 2013 (black points) \cite{ThePierreAuger:2013eja,Aab:2014aea,Aab:2014kda}. Apart from the last energy bin in the RMS data, we can conclude that a mixture of astrophysical sources and a component produced by the decay of SHDM is compatible with the chemical composition observed by Auger. The chemical composition expected with different choices of $\beta$ remains practically unchanged, therefore in figure \ref{fig2} we have plotted only the case $\beta=2$. 

Future observatories of UHECRs and neutrinos should be able to discover SHDM or constrain their lifetimes. For UHECRs, the Extreme Universe Space Observatory to be deployed in the Japanese Experiment Module of the International Space Station (JEM-EUSO) should achieve about an order of magnitude higher exposure at the $10^{20}$ eV energy region as compared to current  ground arrays \cite{Ebisuzaki:2014wka}. A possible future network of such space observatories could continue to improve sensitivity up to the observational limit of the Earth's atmosphere, which is another four orders of magnitudes in exposure. For UHE neutrinos, the upcoming Askaryan Radio Array (ARA) \cite{Allison:2014kha,Allison:2011wk} will improve by about an order of magnitude the sensitivity to $10^{20}$ eV when compared to the strongest current limits of ANITA II \cite{Gorham:2010kv,Gorham:2008yk}.

\begin{figure}
   \centering
   \includegraphics[width=0.495\textwidth]{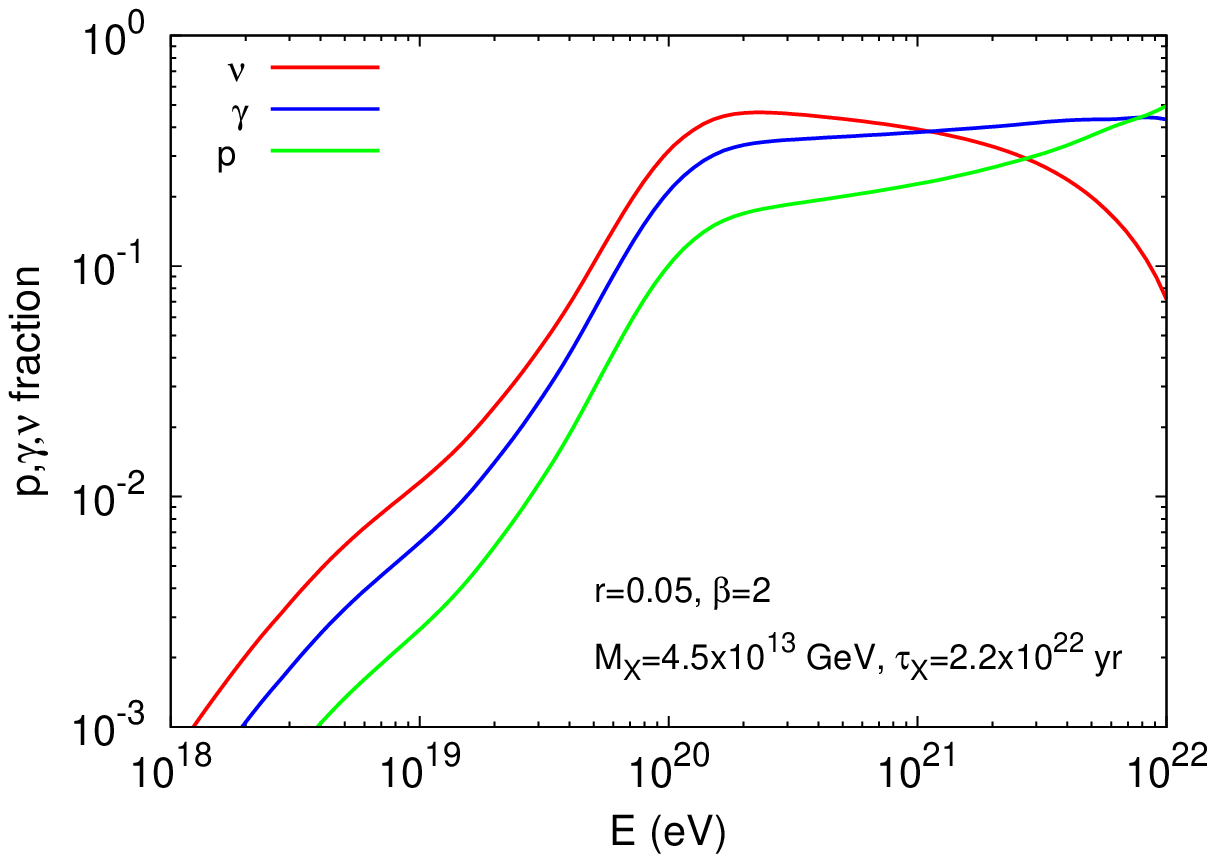}
   \includegraphics[width=0.495\textwidth]{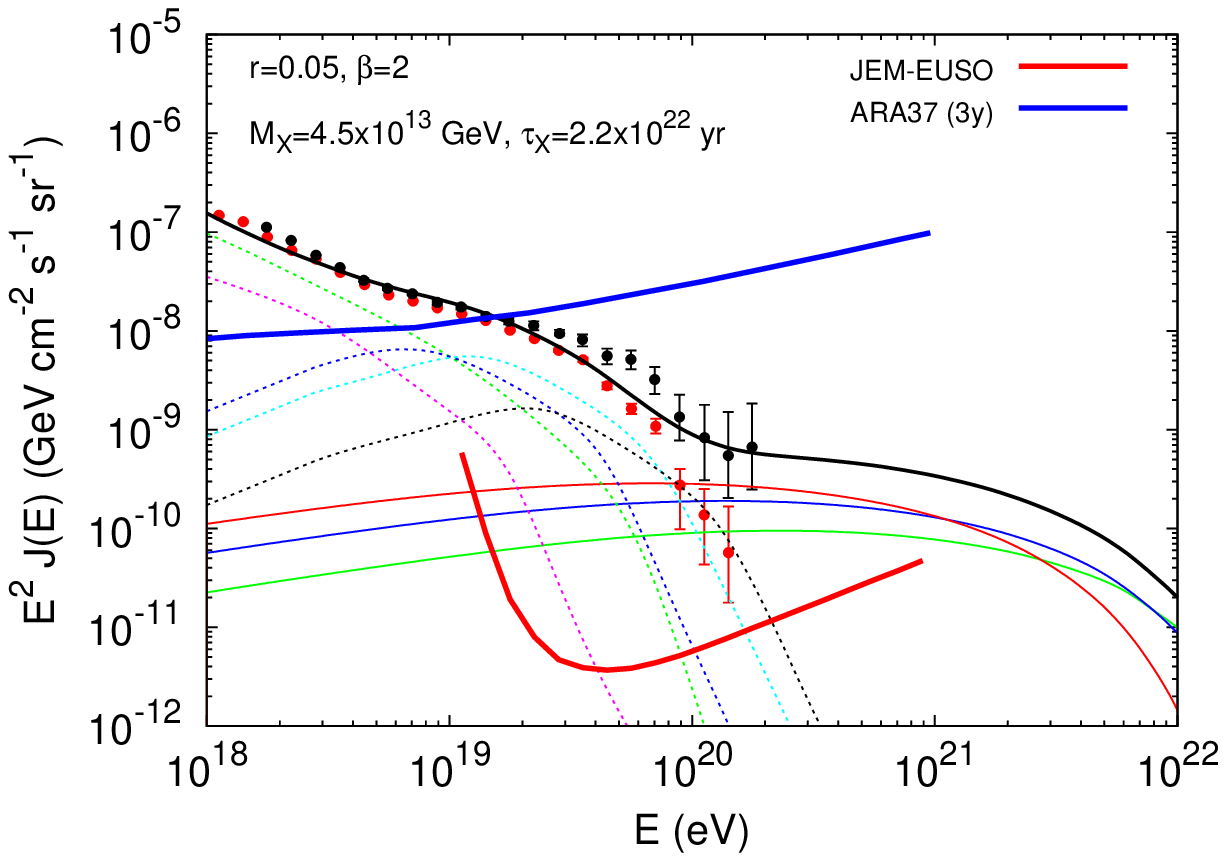}      
   \caption{[Left panel] Fraction over the total UHECR flux of protons, photons and neutrinos by SHDM decay, in the case of $r=0.05$ and $\beta=2$. [Right panel] Sensitivity to SHDM decay products: for UHECRs the future JEM-EUSO space mission (thick red solid line) and for UHE neutrinos the upcoming ARA observatory (thick blue solid line). Experimental data are those of Auger (red points) \cite{ThePierreAuger:2013eja} and TA (black points) \cite{Abu-Zayyad:2013qwa}. Theoretical fluxes are the same of upper left panel in figure \ref{fig1} ($r=0.05$, $\beta=2$).}
   \label{fig3} 
\end{figure}

The most striking signature of SHDM models is represented by the peculiar composition of UHECR at the highest energies, with a flux dominated by neutrinos and photons, which is an exact outcome of the sole decaying dynamics of the SHDM particles \cite{Aloisio:2003xj}. In figure \ref{fig3} (left panel) we plot the fraction of protons (only those produced by SHDM), photons and neutrinos over the total flux (taking again the case $\beta=2$, $r=0.05$ in both panels of the figure).

In figure \ref{fig3} (right panel) we plot the flux of UHE cosmic rays, gamma-rays, and neutrinos from SHDM decay and the sensitivity of ARA to neutrinos (thick blue solid line) and JEM-EUSO (thick red solid line) to cosmic and gamma-rays. Right panel of figure \ref{fig3} shows how neutrino observatories are not competitive with UHECR observatories in the search for SHDM. In the near future, only the JEM-EUSO experiment should be able to detect these secondaries and test the SHDM hypothesis if the predicted ratio of protons, neutrinos, and photons is observed (figure \ref{fig3} left panel). If none of the secondaries are found, a lower limit on the lifetime of SHDM particles can be derived. 

\begin{figure}
   \centering
   \includegraphics[width=0.495\textwidth]{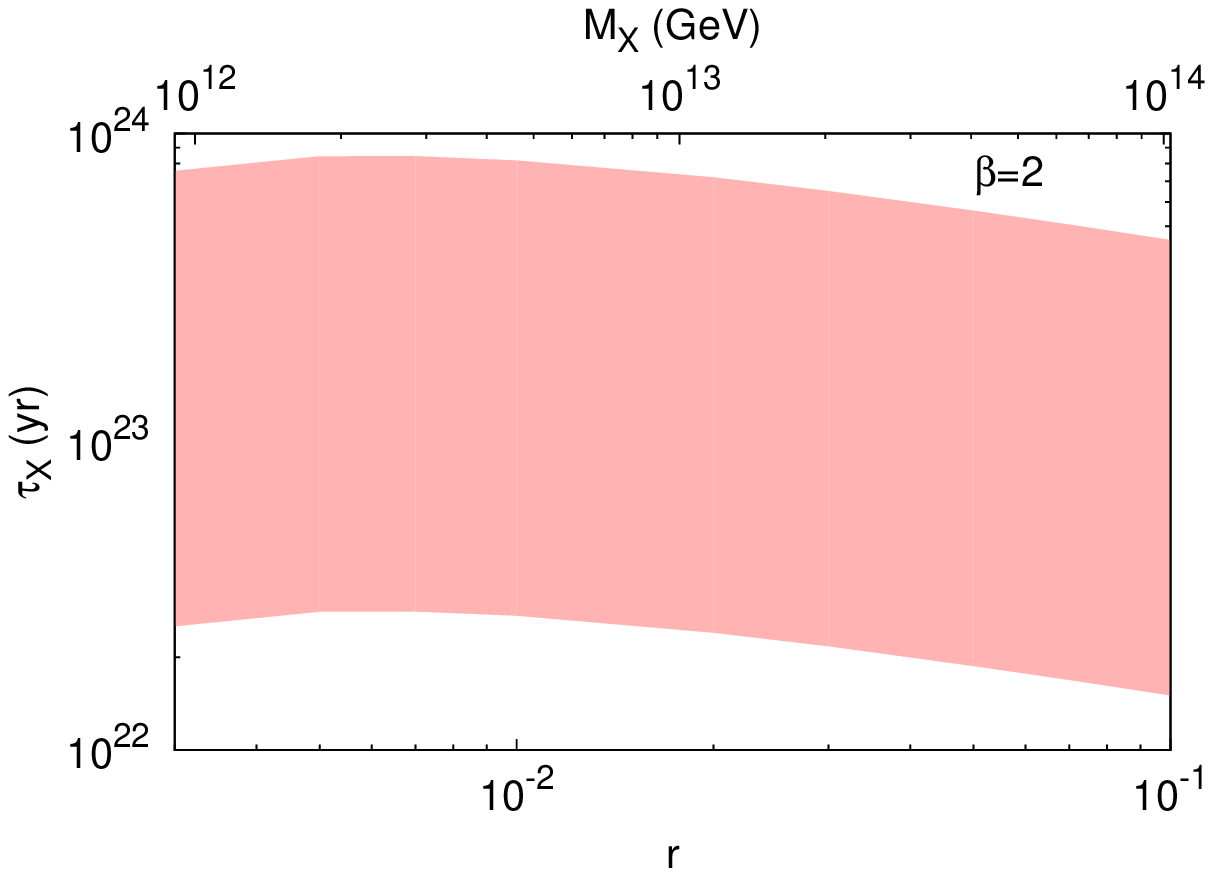}
   \includegraphics[width=0.495\textwidth]{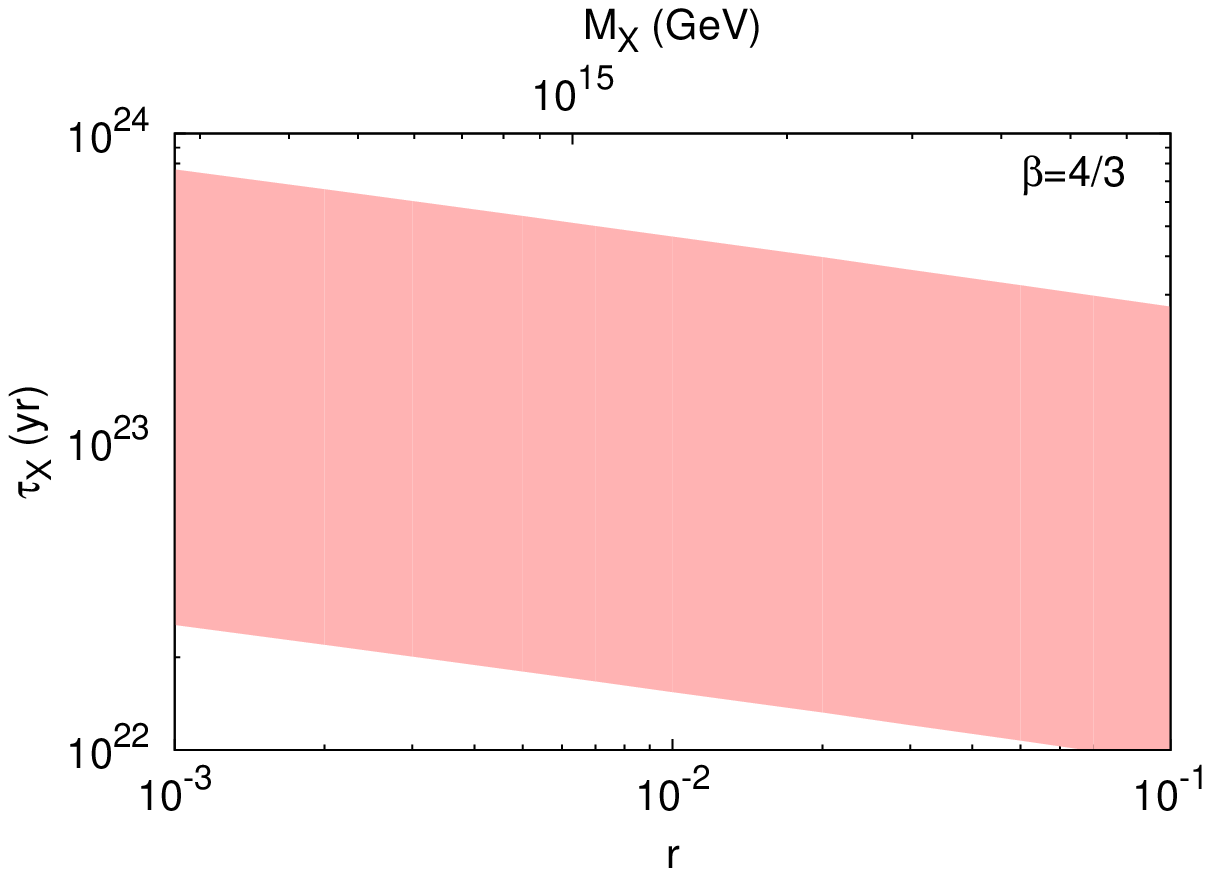}
   \includegraphics[width=0.495\textwidth]{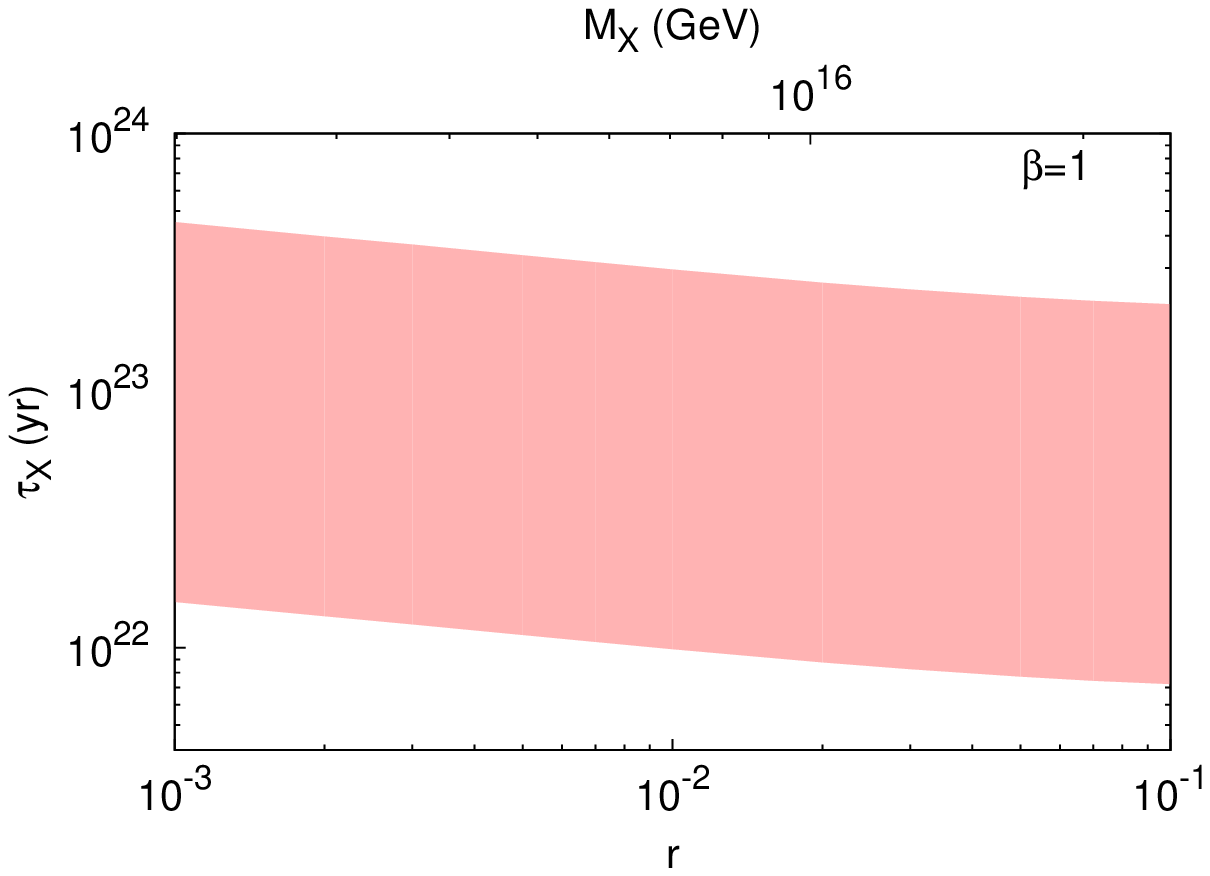}
   \includegraphics[width=0.495\textwidth]{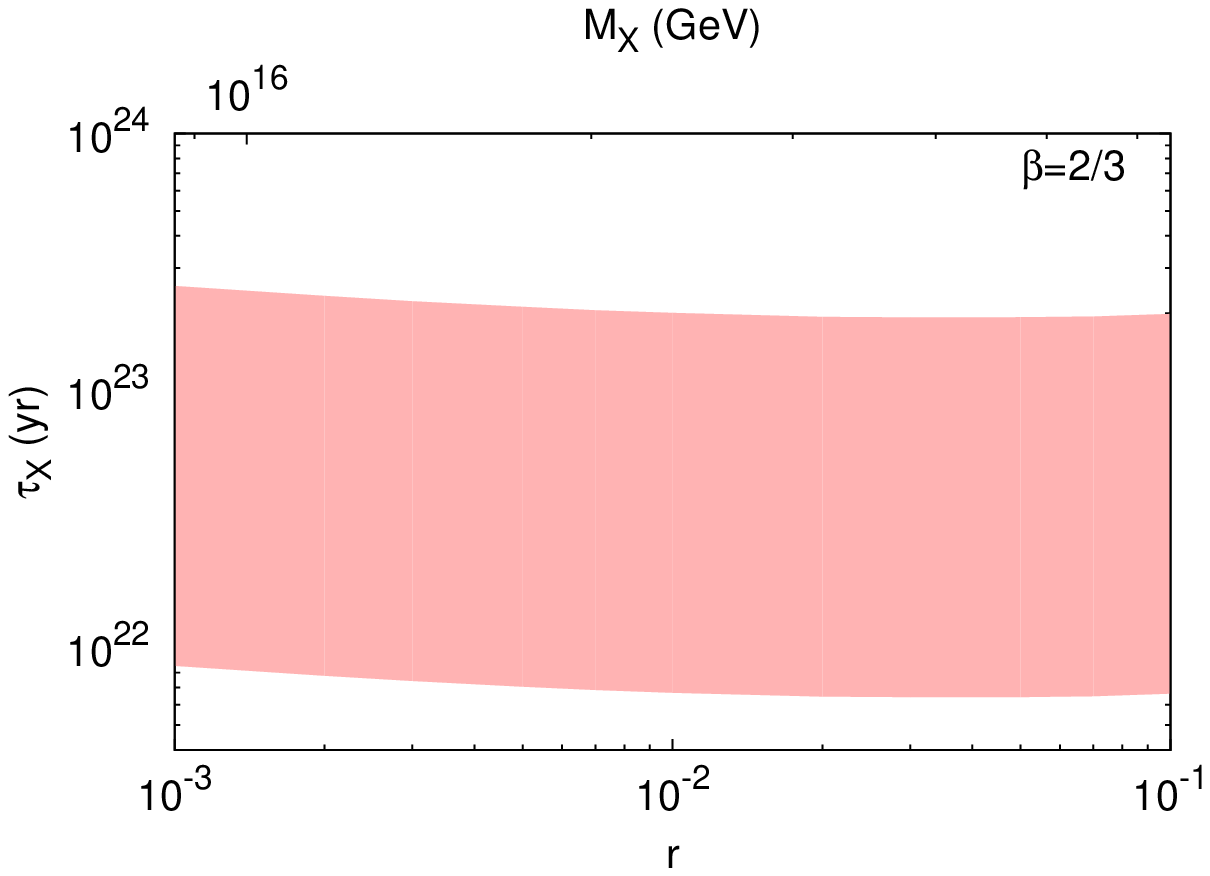}
   \caption{ Shadowed red areas represent the regions in the plane $(r,\tau_X)$ (or $(M_X,\tau_X)$ as labelled in the upper abscissa of each panel) accessible to the JEM-EUSO experiment, each panel corresponds to a different choice of the power law index of the inflaton potential (as labelled): upper left panel $\beta=2$, upper right panel $\beta=4/3$, lower left panel $\beta=1$ and lower right panel $\beta=2/3$.  } 
   \label{fig4} 
\end{figure}

In order to better quantify the JEM-EUSO capabilities in exploring SHDM models, in figure \ref{fig4} we plot the regions on the plane $(r,\tau_X)$ accessible to this experiment. While the SHDM mass $M_X$ is univocally fixed by $r$ (and the inflaton potential), to determine the detectable values of $\tau_X$ we assumed a JEM-EUSO exposure that enables the observation of UHECR fluxes from SHDM a factor of $30$ lower than the value determined by the Auger limits on $\gamma$ rays at $E\ge 10^{19}$ eV and on the total number of events observed by Auger at $E\ge 6\times 10^{19}$ eV. The latter being the minimum allowed value of $\tau_X$ used in figures \ref{fig1}, \ref{fig2} and \ref{fig3}. In figure \ref{fig4} we have considered the tensor to scalar ratio in the range $(10^{-3},10^{-1})$, as follows from the latest combined analysis of BICEP2, Planck and Keck Array \cite{Ade:2015tva}. Only in the case $\beta=2$ we restricted our computations to the interval $3\times 10^{-3}<r<10^{-1}$ because of our assumptions on the reheating temperature as discussed in the previous section (see right panel of figure \ref{fig0}). The four panels of figure \ref{fig4} refer to the four different choices of the inflaton potential power law index $\beta=2/3,1,4/3,2$ (as labelled in the figure). In each panel the lower abscissa represents the tensor to scale ratio $r$ while the upper one shows the corresponding value of the SHDM mass $M_X$ that assures $\Omega_X=\Omega_{DM}$ today. 

\section{Conclusions} 
\label{Conclu}

The BICEP program represents great progress in the search for primordial B-mode polarization. This encouraging trend revived the notion that the signal may be higher than previously expected. However, Plack showed that subtracting foreground contamination by dust is also more challenging than past estimates. The joint Planck, BICEP2, and Keck Array analysis \cite{Ade:2015tva} set an upper limit on $r$ while mildly pointing towards a non-zero possibility. 
Given the large interest in the community and the ability of next generation experiments to surpass current challenges, we are optimistic that  the fundamental measurement of $r$ will be reached in the near future, if it is not too defiantly small. 

Relatively large values of $r$, which may be detected in the near future, motivate the idea that dark matter is mainly composed of superheavy relics from the inflationary epoch. Given a measurement of $r$,  the existence of SHDM can best be tested by exploring the decay lifetime parameter range with future UHECR observatories. The higher the statistics of UHECR experiments at energies around $10^{20}$ eV, the more likely the detection of SM particles produced by SHDM decay. Given the currently planned UHECR and UHE neutrino detectors, JEM-EUSO is best placed to effectively study the allowed SHDM lifetimes with possible lifetime detections or constraints reaching values as high as $\tau_X\simeq 10^{24}$ yr.

\section*{Acknowledgments}
The authors thank D. Chung, P. Lipari, M. Raidal, F. Salamida  and F. Vissani for fruitful discussions about BICEP2 observations, SHDM hypothesis and UHECR observations. The authors would like to thank also the anonymous referee who helped improve the paper. RA and SM thank the Gran Sasso Science Institute where the initial idea of this work was developed. AVO acknowledges the support of NASA award NNX13AH54G, NSF grant PHY-1412261 and the KICP through grant NSF PHY-1125897 and the Kavli Foundation at the University of Chicago.

\bibliographystyle{JHEP} 
\bibliography{SHDMbib}

\end{document}